\documentclass[12pt]{article}
\usepackage{epsfig}
\renewcommand{\baselinestretch}{1.2}
\makeatletter
\renewcommand{\section}{\setcounter{equation}{0}\@startsection
  {section}%
  {1}%
  {0pt}%
  {-1\baselineskip}%
  {0.4\baselineskip}%
  {\large \bfseries}}%
\renewcommand{\subsection}{\@startsection
  {subsection}%
  {2}%
  {0pt}%
  {-0.75\baselineskip}%
  {0.2\baselineskip}%
  {\bfseries}}%
\renewcommand{\subsubsection}{\@startsection
  {subsubsection}%
  {3}%
  {0pt}%
  {-0.5\baselineskip}%
  {0.1\baselineskip}%
  {\sc}}%
\makeatother

\renewenvironment{itemize}
  {\begin{list}%
     {}%
     {\setlength{\topsep}{-6pt}%
      \setlength{\partopsep}{-6pt}%
      \setlength{\itemsep}{-3pt}%
      \setlength{\labelsep}{5pt}%
      \setlength{\itemindent}{0pt}%
     }%
  }%
  {\end{list}}%

\def\cD{{\cal D}}
\def\cR{{\cal R}}

\def\a{\alpha} 
\def\b{\beta}
\def\d{\delta}          
\def\De{\Delta}
\def\ga{\gamma}         
\def\gm{\Gamma}
 
\def\la{\lambda}        
\def\La{\Lambda}
\def\ka{\kappa}
\def\m{\mu}
\def\n{\nu}
\def\r{\rho}
\def\s{\sigma}
\def\t{\tau}
\def\th{\theta}
\def\eps{\epsilon}
\def\ee{\varepsilon}
\def\om{\omega}         
\def\Om{\Omega}

\def\oom{\bar\omega}
\def\uom{\underline\omega}

\def\parslash{{\partial{\hspace{-6pt}}/\hspace{1pt}}}
\def\Aslash{{A\hspace{-7pt}/\hspace{2pt}}}
\def\Vslash{{V\hspace{-9pt}/}\hspace{5pt}}
\def\Dslash{{D\hspace{-7pt}/\hspace{2pt}}}
\def\Kslash{{K\hspace{-9pt}/\hspace{4pt}}}
\def\qslash{{q\hspace{-5pt}/}}
\def\kslash{{k\hspace{-5.5pt}/\hspace{0.5pt}}}

\def\tg{\tilde{g}}
\def\hg{\hat{g}}
\def\tp{\tilde{p}}
\def\hp{\hat{p}}

\def\YM{\rm YM}
\def\CS{\rm CS}
\def\GF{\rm GF}
\def\ES{\rm ES}

\def\idp{\int\! {d^3\!p \over (2\pi)^3} \,\,}
\def\iddq{\int\! {d^d\!q \over (2\pi)^d} \,\,}
\def\idx{\int\! d^3\!x \,}
\def\idxth{\int\! d^3\!x \,d^2\!\th \,}

\def\ds{\displaystyle}
\def\ss{\scriptstyle}
\def\to{\rightarrow}
\def\igual{\hspace{-7pt}=\hspace{-7pt}}
\def\mas{\hspace{-7pt}+\hspace{-7pt}}
\def\menos{\hspace{-7pt}-\hspace{-7pt}}
\def\RR{{\rm I\!\!\, R}}
\def\unit{{\rm 1\hspace{-2pt}l}}

\begin{document}
\begin{titlepage}
\rightline{HD-THEP-96-29}
\rightline{ITP-SB-96-36}
\rightline{NIKHEF-96-019} 
\vskip 2 true cm
\begin{center}
{\Large {\bf BRS symmetry versus supersymmetry in }}\\ 
\vskip 0.3 true cm 
{\Large {\bf Yang-Mills-Chern-Simons theory}}\\ 
\vskip 1.2 true cm 
{\rm F. Ruiz Ruiz}\footnote{E-mail: ruiz@thphys.uni-heidelberg.de. 
     Alexander von Humboldt Research Fellow. Research partially 
     supported by CICyT grant AEN95-1284E}\\ 
\vskip 0.3 true cm 
{\it Institut f\"ur Theoretische Physik, Universit\"at Heidelberg}\\
{\it Philosophenweg 16, 69120 Heidelberg, Germany and}\\ 
{\it NIKHEF, Postbus 41882, 1009 DB Amsterdam, The Netherlands}\\ 
\vskip 1 true cm 
{\rm P. van Nieuwenhuizen}\footnote{E-mail: 
     vannieu@insti.physics.sunysb.edu. Research supported by NSF 
     grant Phy 9309888.}\\ 
\vskip 0.3 true cm
{\it Institute for Theoretical Physics, State University of New York
     at Stony Brook}\\ 
{\it Stony Brook, NY 11794-3840, USA}\\ 

\vskip 1.5 true cm

{\leftskip=1.5 true cm \rightskip=1.5 true cm 
\noindent
We prove that three-dimensional $N=1$ supersymmetric
Yang-Mills-Chern-Simons theory is finite to all loop orders. In
general this leaves open the possibility that different regularization
methods lead to different finite effective actions.  We show that in
this model dimensional regularization and regularization by
dimensional reduction yield the same effective action.  Consequently,
the superfield approach preserves BRS invariance for this model.
\noindent
\par
}
\end{center}
\end{titlepage}
\setcounter{page}{2}


\section{Introduction and conclusions}

One of the major unsolved problems in supersymmetry is the
supersymmetric regularization of gauge theories. The renormalized
effective action that results from using a particular regularization
method and subtraction prescription can be made to satisfy either the
supersymmetry or the gauge Ward identities by adding suitable finite
local counterterms, but then the question arises whether the new
renormalized effective action satisfies the Ward identities of the
other symmetry. On algebraic cohomological grounds, it has been argued
that for certain supersymmetric theories there exist renormalized
effective actions which satisfy both sets of identities \cite{Piguet}.
However, this does not tell us how to actually compute such preferred
renormalized actions. In fact, no regularization method for
four-dimensional supersymmetric gauge theories preserving both gauge
invariance and supersymmetry is known to date.

In this article we study the formulation of a supersymmetric and gauge
invariant regularization method in three dimensions. We consider
$N\!=\!1$ supersymmetric Yang-Mills-Chern-Simons theory, whose
classical action 
\begin{displaymath}
   S = {1\over m}\, S_{\YM} + S_{\CS} 
\end{displaymath}
consists of the sum of the Yang-Mills and the Chern-Simons actions,
and study two regularization methods, ordinary dimensional
regularization (or DReG) and regularization by dimensional reduction
(or DReD). We are interested in the difference $\De\gm=\gm^{\rm DReG}
[\psi,K_\phi] - \gm^{\rm DReD}[\psi,K_\phi]$ of the corresponding
effective actions, where $\psi$ stands for all the fields and $K_\phi$
for the sources of the fields with nonlinear BRS transformations.
Since we will show that the theory is finite, the regularized
effective actions $\gm^{\rm DReG}$ and $\gm^{\rm DReD}$ are also
renormalized effective actions and $\De\gm$ is the difference of two
renormalized effective actions. The first regularization method,
namely DReG \cite{tHooft}, preserves at all stages the BRS identities
corresponding to local gauge invariance. This is so since, by treating
the $\eps\!$-symbol $\eps^{\m\n\r}$ in the classical Chern-Simons
action as purely three-dimensional \cite{GMR}, the kinetic matrix for
the gauge field has an inverse in $d\!\geq\!3$ dimensions and, by
using this inverse as gauge propagator, the BRS symmetry is maintained
in $d$ dimensions.  Unfortunately, because for $d\!\neq\! 3$ the
numbers of bosons and fermions are not equal, even when the Dirac
algebra of the Feynman diagrams is performed in $d$ dimensions, DReG
does not preserve supersymmetry manifestly. The second method,
regularization by DReD \cite{Siegel-DReD}, performs the algebra of all
Feynman diagrams in terms of superfields and, only at the end,
continues the momentum integrals to $d\!<\!3$ dimensions.  The
propagator that DReD uses for the gauge field is not the inverse of
the kinetic term in $d$ dimensions because such an inverse does not
exist for $d\!<\!3.$ For this reason, DReD does not preserve BRS
invariance at all stages and runs the risk of violating the BRS
identities, although it preserves supersymmetry manifestly. The
well-known inconsistency of DReD \cite{Siegel-incon} does not occur
for this model (see section 3). 

Our strategy will be:
\begin{itemize}
\item[(i)] To show that the theory is finite to all loops. In fact, by
  power counting the theory is only superrenormalizable, since it
  contains divergences at the one, two and three-loop levels. Hence,
  to compute radiative corrections, regularization is needed. We will
  use dimensional regularization to prove finiteness.
\item[(ii)] To use this to prove that the difference $\De\gm$ is
proportional to a supersymmetric polynomial of the fields and their
derivatives with only one free parameter. 
\item[(iii)] To compute this parameter and to show that it vanishes.
\end{itemize}
Since the difference $\De\gm$ vanishes, one may use superspace methods
to compute loop corrections while preserving both BRS invariance and
supersymmetry. Our results depend critically on the
fact that we are in three dimensions, and we make no claims concerning
four-dimensional theories.

We will prove finiteness at one loop by using properties of
dimensionally regularized integrals. At two loops, finiteness follows
from the observation that there are no one-loop subdivergences and
that the two-loop effective action $\gm^{\rm DReG}_2$ satisfies the
BRS identity
\begin{equation}
  \Theta \gm^{\rm DReG}_2 + (\gm^{\rm DReG}_1,\gm^{\rm DReG}_1) = 0~,
\label{BRS-intro}
\end{equation}
so that the divergent part satisfies $\Theta \gm^{\rm DReG}_{\rm
  2,div} = 0,$ where $\Theta$ is the Slavnov-Taylor operator. Since
the divergences in a 1PI Green function at $k$ loops are polynomials
in the external momenta with degree equal to or less than the
superficial overall UV degree of divergence $\oom_k$ of the
corresponding proper graphs, the most general form of $\gm^{\rm
  DReG}_{\rm 2,div}$ is $\gm^{\rm DReG}_{\rm 2,div}={1\over
  d-3}\,P_{\oom_2}[\partial,\psi,K_\phi],$ with $P_{\oom_2}$ a certain
polynomial in the fields, sources and their derivatives. To determine
which terms are possible in $P_{\oom_2},$ we need power counting for
the various 1PI diagrams. We will find that no 1PI diagrams with BRS
sources are superficially divergent, but only 1PI diagrams with
fields. At one loop we will find quadratic, linear and logarithmic
divergences, while at two loops only linear and logarithmic, and at
three loops only logarithmic divergences will remain. In particular,
$P_{\oom_2}$ will only depend on the component fields in the gauge
multiplet. We will show that no BRS invariant can be constructed out
of these components, so that $\gm^{\rm DReG}_{\rm 2, div}=0.$ The same
arguments will prove that the theory is also finite at three and
higher loops.

There is a general theorem in quantum field theory \cite{Hepp}
\cite{Maison} that states that if two different {\it renormalization}
(not regularization) schemes yield the same Green functions up to
$k\!-\!1$ loops, then at $k$ loops they give Green functions that can
differ at most by a local finite polynomial in the external momenta of
degree equal to the superficial overall UV degree of divergence
$\oom_k$ at $k$ loops. To go from the regularized to the renormalized
Green function at $k$ loops, one must in general subtract the
$k\!$-loop divergences.  Given that in our case the theory is finite,
and provided the regularized DReG and DReD expressions for the Green
functions are identical at $k\!-\! 1$ loops, it follows that at $k$
loops they can differ at most by a local finite polynomial in the
external momenta. Using properties of dimensionally regularized
integrals, we will show that DReG and DReD give the same expressions
for all Green functions at one loop, so $\De\gm_1=0.$ Then, the
difference $\De\gm_2$ at two loops will be 
\begin{equation}
  \gm^{\rm DReG}_2 - \gm^{\rm DReD}_2 
        = P_{\oom_2}[\partial,\psi,K_\phi] ~,
\label{diff-2}
\end{equation}
with the same polynomial $P_{\oom_2}$ that comes out in the analysis
of finiteness.  We already know that $\gm^{\rm DReG}_2$ satisfies the
BRS identity (\ref{BRS-intro}). As for $\gm^{\rm DReD}_2,$ since DReD
manifestly preserves supersymmetry, it satisfies the supersymmetry
Ward identity
\begin{equation} 
  \d \gm^{\rm DReD}_2 = 0 \>,
\label{susy-two}
\end{equation}
where $\d$ is the supersymmetry generator. Acting with $\d$ on eq.
(\ref{BRS-intro}), using eqs. (\ref{diff-2}) and (\ref{susy-two}),
noting that $[\Theta,\d]=0$ and recalling that $\De\gm_1=0$ leads to
$\Theta\d P_{\oom_2} = 0$
\begin{displaymath}
  0 = \d\Theta\,\Big(\,\gm^{\rm DReD}_2 + P_{\oom_2}\,\Big) 
    + \d\,(\,\gm^{\rm DReD}_1, \gm^{\rm DReD}_1\,\Big) 
    = \Theta \d P_{\oom_2} \>.
\end{displaymath}
As already mentioned, power counting implies that the terms allowed in
$P_{\oom_2}$ depend only on the component fields of the gauge
multiplet.  Hence, $\d P_{\oom_2}$ is a finite local functional of
such fields. We will show (i) that it is not possible to construct a
BRS invariant from these ingredients, so that $\d P_{\oom_2}=0,$ and
(ii) that the only supersymmetric invariant one can construct from
$P_{\oom_2}$ is
\begin{equation}
   P_{\oom_2}^{\rm susy} = \a\> m \idxth \, \gm^{a\a}\gm^a_\a \> ,
\label{susy-invariant-super}
\end{equation}
with $\gm^a_\a$ the basic spinor superfield containing the gauge
multiplet and $\a$ a constant. Finally, we compute $\a$ and obtain
zero. The same arguments as at two loops imply that the difference
$\De\gm_3$ also vanishes at three and higher loops, except that in
this case there is not even a derivative in $P_{\oom_3}$ so that no
supersymmetric invariant can be constructed. Hence, $P_{\oom_3}=0.$

The paper is organized as follows. In section 2, we introduce our
notation and write the classical action and the BRS and supersymmetry
transformations in superfields and components. We need the component
expressions since ordinary DReG can not be formulated in terms of
superfields. To have a supersymmetric gauge-fixed classical action we
will work in the supersymmetric Landau gauge, which although linearly
realized in terms of the superfield $\gm^a_\a$ will give nonlinear
gauge conditions for the components of the gauge multiplet. In section
3 we define DReG and explain how to handle the $\eps^{\m\n\r}$ in the
classical action. We also briefly recall some elements of DReD.
Section 4 contains our proof of perturbative finiteness, and section 5
the proof that DReG and DReD yield the same expression for the
effective action.  Section 6 contains some further remarks.

\section{Classical action, gauge-fixing and power counting}

We will work in Minkowski spacetime with metric $g_{\m\n}={\rm
  diag}\,(-,+,+)$ and completely antisymmetric tensor $\eps^{\m\n\r}$
defined by $\eps^{012}=1.$ We recall that in three dimensions the
Lorentz algebra can be realized as $so(1,2)$ or as $sl(2,\RR),$ and that
the fundamental representation has dimension two and acts on real
Majorana spinors $\psi^\a.$ Spinor indices will be denoted by Greek
letters $\a,\b,\ldots$ and will be raised and lowered with the
rank-two antisymmetric tensors $\eps^{\a\b}$ and $\eps_{\a\b}$ and
the northwest-southeast convention. That is, $\psi^\a=
\eps^{\a\b}\psi_\b$ and $\psi_\a= \psi^\b\eps_{\b\a}.$ We define
$\eps_{\a\b}$ by $\eps_{12}=1,$ which together with our contraction
convention implies that $\eps^{12}= 1.$ In this paper we will consider
$N\!=\!1$ supersymmetry, so all our spinors will be Majorana. The
vector representation of the Lorentz group has dimension three and
acts on real vectors. A vector $v$ admits an $so(1,2)$ realization as
a spacetime vector $v^\m$ and an $sl(2,\RR)$ realization as a
symmetric rank-two tensor $v^\a_{~\b}.$ To go from one to the other
we use the Dirac gamma matrices $\ga_\m:$
\begin{displaymath}
  v^\a_{~\b} = \big( \ga_\m \big)^\a_{~\b}\, v^\m \,.
\end{displaymath}
The Dirac gamma matrices $\ga^\m$ satisfy $\{\ga^\m,\ga^\n\} = 2
g^{\m\n}$ and have spinor indices $\big(\ga^\m)^\a_{~\b}.$ When
necessary, we will use the real representation $\ga^0=-i\s^2,
~\ga^1=\s^1, ~\ga^2=\s^3.$ The charge conjugate of a spinor $\psi^\a$
is defined by $\bar{\psi}_\b= \psi^\a C_{\a\b},$ with $C_{\a\b}=
-i\eps_{\a\b}$ the charge conjugation matrix. By definition, $C$
satisfies $C=-C^T$ and $(C\ga^\m)^T = C\ga^\m.$ Finally, we recall
that in three dimensions the matrices $\{\unit,\ga_\m\}$ form a basis
of the Clifford algebra and that given four spinors
$\psi_i~(i=1,2,3,4)$ the Fierz identity takes the form
\begin{displaymath}
   (\bar\psi_1 M \psi_2)\, (\bar\psi_3 N \psi_4) = - \,
      {1\over 2}~ \Big[\, 
         (\bar\psi_1 M N \psi_4)\, (\bar\psi_3 \psi_2)   
       + (\bar\psi_1 M\ga^\m N \psi_4)\, (\bar\psi_3 \ga_\m \psi_2)
         \Big] ~.
\end{displaymath}

Our conventions for superfields are as follows.  Superspace is
parameterized by three real spacetime coordinates $x^{\a\b}$ and two
real anticommuting Majorana spinor coordinates $\th^\a.$ Spacetime
derivatives are denoted by $\partial_{\a\b} \equiv \big( \ga^\m
\big)_{\a\b}\,\partial_\m,$ ordinary spinor derivatives by
$\partial_\a$ and spinor superderivatives by $D_\a = \partial_\a\!  +
i\,\th^\b\partial_{\b\a}.$ Useful identities to project onto
components are $\{D_\a,D_\b\}= 2i\partial_{\a\b},~
[D_\a,D_\b]=-\eps_{\a\b}D^2$ and $D^\b D_\a D_\b=0.$ From
$\{\partial_\a,\th^\b\}=\d_\a^{~\b}$ and $[\partial_{\a\b},
x^{\ga\d}]= -{1\over 2}\,(\d_\a^{~\ga} \d_\b^{~\d} + \d_\a^{~\d}
\d_\b^{~\ga}),$ it follows that $\partial_\a$ is real and that
$\partial_{\a\b}$ is imaginary, so that $D_\a$ is real. As for the
measure in superspace, we note that $d^3\!x$ is real and
$d^2\!\th\equiv -2\,d\th^1 d\th^2=D^2$ imaginary. Under a
supersymmetry transformation,
\begin{displaymath}
  \d x^{\a\b}\!=\!a^{\a\b}\!\!- 2i\, \ee^{(\a} \th^{\b)} \qquad
  \d\th^a=\ee^\a \,,
\end{displaymath}
where $a^{\a\b}$ is a real commuting constant and $\ee^{(\a}
\th^{\b)}= {1\over 2}\,(\ee^\a \th^\b + \ee^\b\th^\a),$ with $\eps^\a$
an anticommuting constant Majorana spinor. The supercharge $Q_\a$ is
$Q_\a =\partial_\a\! - i\,\th^\b\partial_{\b\a}.$ As usual, a
superfield $\Psi(x,\th)$ transforms linearly under the action of the
supercharge: $\d \Psi = \eps^\a Q_\a\Psi.$

\subsection{Superfield classical action}

We assume the gauge algebra to be a real, compact, semi-simple Lie
algebra, so that the structure constants $f^{abc}$ can be taken
completely anti-symmetric without loss of generality. The real gauge
field $A^a_{\a\b}$ is part of a vector supermultiplet described by a
Majorana spinor gauge potential $\gm^a_\a$ \cite{1001}.
\begin{table}[ht]
\begin{center}
\begin{tabular}{|l|c|c|c|c|c|c|c|c|c|c|c|c|c|}
\hline
  field & $\chi^a_\a$ & $H^a$ & $A^a_{\a\b}$ & $\la^a_\a$ & $b^a$ 
  & $\zeta^a_\a$  & $h^a$ & $c^a$ & $\varphi^a_\a$ & $\om^a$ 
  & $\hat{c}^a$ & $\hat{\varphi}^a_\a$ & $\hat{\om}^a$ \\
\hline
  mass dim. & 1/2 & 1 & 1 & 3/2 & 1 & 3/2 & 2 & 1/2 & 1 & 3/2
  & 1/2 & 1 & 3/2  \\
\hline
  ghost no. & 0 & 0 & 0 & 0 & 0 & 0 & 0 & 1 & 1 & 1 & -1 & -1 & -1 \\  
\hline
  Grassmann grad. & A & C & C & A & C & A & C & A & C & A & A & C & A
\\ \hline
\end{tabular} 
\\[9pt]
\renewcommand{\baselinestretch}{0.8}
{\leftskip=2.7 true cm \rightskip=2.7 true cm \sl 
Table 1: mass dimension, ghost number and Grassmann grading
(A=anticommuting, C=commuting) of component fields.
\par}
\end{center}
\vspace{-15pt}
\end{table}
\renewcommand{\baselinestretch}{1.2} Besides $A^a_{\a\b},$ the
supermultiplet contains a real scalar field $H^a$ and two
anticommuting Majorana spinors $\chi^a_\a$ and $\la^a_\a.$ The
superfield $\gm^a_\a$ defines a real vector gauge potential
$\gm^a_{\a\b}$ and an imaginary spinor field strength $W^a_\a$ through
the expressions
\begin{displaymath}
\begin{array}{c} {\ds
  \gm^a_{\a\b} = D_{(\a} \gm^a_{\b)} 
     + {i\over 2}\> f^{abc} \,\gm^b_\a \gm^c_\b }\\[9pt]
  {\ds W^a_\a = D^\b D_\a \gm^a_\b + i f^{abc} \, \gm^{b\b} D_\b\gm^c_\a
    - {1\over 3}\, f^{abc} f^{cde}\,\gm^{b\b}\,\gm^d_\b\, \gm^e_\a ~.}
\end{array}
\end{displaymath}
The reality of $\gm^a_{\a\b}$ follows from writing $D_\a\gm^a_\b$ as
$\{D_\a,\gm^a_\b\}$ and using that $D_\a$ is real. As a check on the
coefficients in $W_\a^a,$ one may verify that the Bianchi identity
$\nabla^\a W^a_\a=0$ is satisfied. In terms of $\gm^a_\a,~
\gm^a_{\a\b}$ and $W^a_\a,$ the components of the gauge multiplet are
given by
\begin{equation}
  \chi^a_\a = \gm^a_\a \Big\vert \qquad
  H^a = {1\over 2}\, D^\a \gm^a_\a \Big\vert \qquad
  A^a_{\a\b} = \gm^a_{\a\b} \Big\vert \qquad
  \la^a_\a = -\, {i\over 2}\, W^a_\a \Big\vert ~,
\label{gauge-comp}
\end{equation} 
where the vertical bar denotes projection onto $\th^\a\!=\!0$ and the
numerical factors have been adjusted so that the Yang-Mills and
Chern-Simons component actions have the standard form [see eqs.
(\ref{YM-comp})-(\ref{CS-comp})]. We take $\gm^a_\a$ to have mass
dimension 1/2, so that using that $\theta^\a$ has mass dimension
$-1/2,$ the components have mass dimensions as in Table 1.

The classical $N\!=\!1$ Yang-Mills-Chern-Simons action has the form
\cite{1001}
\begin{displaymath}
   S = {1\over m} \> S_{\YM} + S_{\CS} ~,
\end{displaymath}
where
\begin{equation}
  S_{\YM} = - {1\over 32\,g^2} \idxth \,W^{a\a} W^a_\a 
\label{YM-super}
\\
\end{equation}
\begin{equation}
 S_{\CS} = {i\over 16\,g^2} \idxth \,\bigg[\,
      \big( D^\a \gm^{a\b}\big) \,\big( D_\b \gm^a_\a\big)
    + {2i\over 3} \,f^{abc}\, 
          \gm^{a\a}\gm^{b\b}\,\big(D_\b\gm^c_\a\big) 
    - {1\over 6}\, f^{abc} f^{cde} \,\gm^{a\a}\,
          \gm^{b\b}\,\gm^d_\a \, \gm^e_\b  \bigg] 
\label{CS-super}
\end{equation} 
are the Yang-Mills and Chern-Simons actions, $m$ is a parameter with
dimensions of mass and $g$ is a dimensionless coupling constant. Both
$S_{\YM}$ and $S_{\CS}$ are invariant under gauge transformations
$\d_\Om \gm^a_\a= \big(\nabla_\a \Om\big)^a,$ where $\nabla^{ab}_\a =
\d^{ab} D_\a + i f^{acb}\, \gm^c_\a$ is the spinor covariant
derivative and $\Om^a$ is an arbitrary real scalar superfield, and
under supersymmetry transformations $\d \gm^a_\a=\ee^\b Q_\b\gm^a_\a.$
To fix the gauge, we impose the supersymmetric Landau condition
\begin{equation}
   D^\a \gm^a_\a = 0 ~.
\label{gauge-super}
\end{equation}
The Faddeev-Popov procedure then adds to the classical action a
contribution
\begin{equation}
  S_{\GF} 
    = {i\over 4} \idxth\, \Big[\, B^a \big(D^\a\gm^a_\a\big) + i\,
    \hat{C}^a \big(D^\a \nabla_\a C\big)^a \Big] ~,
\label{GF-super}
\end{equation}
with $B^a$ a real commuting Lagrange multiplier superfield of mass
dimension 1 imposing the condition $D^\a\gm^a_\a\!=\!0,$ and
$\hat{C}^a$ and $C^a$ real anticommuting antighost and ghost
superfields of mass dimension 1/2. After gauge fixing, gauge
invariance is replaced by BRS invariance. To obtain the BRS variation,
we replace the gauge parameter $\Om^a$ with $-i\eta C^a$ and $\d_\Om$
with $\eta s,$ where $\eta$ is a Grassmann constant, $C^a$ is the
ghost superfield and $s$ is the BRS operator. This and the requirement
of nilpotency for $s$ gives the following BRS transformations:
\begin{equation}
   s\,\gm^a_\a = i\big( \nabla_\a C\big)^a \qquad
   s\,B^a =0 \qquad 
   s\,\hat{C}^a = B^a \qquad 
   s\,C^a= -{1\over 2} \,f^{abc}\, C^b\, C^c \,.
\label{BRS-super}
\end{equation}
It can easily be checked that $s$ leaves $S_{\YM},~S_{\CS}$ and
$S_{\GF}$ invariant. We note that $[s,\d]=0$ because they act in
different spaces. We define the field components of $B^a,~C^a$ and
$\hat{C}^a$ through the projections (see Table 1 for their mass
dimension and Grassmann grading)
\begin{equation} 
\begin{array}{rrrl}
     b^a = B^a \,\Big\vert 
   & \qquad\qquad\qquad c^a = C^a \,\Big\vert 
   & \qquad\qquad\qquad \hat c^a = \hat C^a\Big\vert &
\\[6pt]
     \zeta^a_\a = i D_\a B^a \,\Big\vert 
   & \varphi^a_\a = D_\a C^a \,\Big\vert 
   & \hat \varphi^a_\a = D_\a \hat C^a\,\Big\vert &
\\[6pt]
     h^a = -\, {\ds {i\over 2}}\, D^2 B^a \,\Big\vert  
   & \om^a = -\, {\ds {i\over 2}}\, D^2 C^a \,\Big\vert  
   & \hat \om^a = -\, {\ds {i\over 2}}\,  D^2 \hat C^a\,\Big\vert 
   & \hspace{-5pt} .
\end{array}
\label{ghost-comp}
\end{equation}
With the purpose of studying BRS invariance at the quantum level, we
introduce commuting external supersources $K^{a\a}_\gm$ and $K^a_C$
coupled to the nonlinear BRS transforms $s\gm^a_\a$ and $sC^a:$
\begin{equation}
   S_{\ES} = {i\over 2} \idxth \,\bigg( \, 
        {1\over 2} \, K^{a\a}_\gm\,s\gm^a_\a - K^a_C\,sC^a \bigg) ~.
\label{ES-super}
\end{equation}
We define the components of $K^{a\a}_\gm$ and $K^a_C$ by the
projections 
\begin{equation} 
\begin{array}{rrl}
     \ka_\a^a = K^a_{\a\,\gm} \,\Big\vert 
   & \qquad\qquad\qquad \ell^a = K_C^a \,\Big\vert & 
\\[6pt]
     {\ds G^a = -\,{i\over 2}\> D^\a K^a_{\a\,\gm} \,\Big\vert }
   & \tau^a_\a = i\,D_\a K_C^a \,\Big\vert &
\\[9pt]
     K^a_{\a\b} = i\,D_{(\a} K^a_\b{}_{)\,\gm} \,\Big\vert 
   & {\ds L^a = -\,{i\over 2}\> D^2 K_C^a \,\Big\vert } &
     \hspace{-5pt} ,
\\[6pt]
     {\ds \s^a_\a = -\, {i\over 2}\, 
         D^\b D_\a K^a_{\b\,\gm} \,\Big\vert}  & 
\end{array}
\label{ext-comp}
\end{equation}
from which the mass dimensions and Grassmann gradings in Table 2
follow.
\begin{table}[ht]
\begin{center}
\begin{tabular}{|l|c|c|c|c|c|c|c|c|c|c|}
\hline 
  external source & $\ka^a_\a$ & $G^a$ & $K^a_{\a\b}$ 
  & $\s^a_\a$  & $\ell^a$ & $\t^a_\a$ & $L^a$ \\
\hline
  mass dim. & 1 & 3/2 & 3/2 & 2 & 1 & 3/2 & 2 \\
\hline 
  ghost no. & -1 & -1 & -1 & -1 & -2 & -2 & -2 \\
\hline
  Grassmann grad. & C & A & A & C & C & A & C \\
\hline
\end{tabular} 
\\[9pt]
{\sl Table 2: Mass dimension, ghost number and Grassmann
grading of external sources.} 
\end{center}
\vspace{-15pt}
\end{table}

All in all, we take as starting point the tree-level action
\begin{equation}
  \gm_0 = {1\over m}\> S_{\YM} + S_{\CS} + S_{\GF} + S_{\ES} ~.
\label{gamma-0}
\end{equation}
Superpower counting for $\gm_0$ gives a finite number of superficially
divergent 1PI diagrams, namely those in Table 3, where $\oom$ denotes
the overall superficial UV degree of divergence. This shows that the
theory is superrenormalizable. We remark that there are no
superficially divergent diagrams with either ghosts $C^a,\,\hat{C}^a$
or sources $K^{a\a}_\gm,\,K^a_C$ as external lines. The connected
generating functional $W[J_\Psi,K_\Phi]$ is given by
\begin{displaymath}
   \exp \Big\{\,i\,W[J_\Psi,K_\Phi]\,\Big\} 
       = \int \! \prod_\Psi \,[d\Psi] \, 
         \exp \Big\{\, i \,\Big( \gm_0[K_\Phi] 
                    + \idxth J_\Psi \Psi \Big)\,\Big\} ~,
\end{displaymath}
where we have introduced external sources $J^a_\Psi=J^{a\a}_\gm,\,
J^a_B,\, J^a_C,\, J^a_{\hat C}$ for the fields $\Psi^a=\gm^a_\a,\,
B^a,\, \hat{C}^a,\, C^a$ and used the notation $K^a_\Phi=
K^{a\a}_\gm,\, K^a_C$ for the sources coupled to the nonlinear BRS
transforms. Performing a BRS change of variables under the integral,
using that there are no BRS anomalies \footnote{Absence of BRS
  anomalies follows from the fact that DReG preserves BRS invariance
  in $n$ dimensions for our model (see section 3). In four-dimensional
  chiral gauge theories, this is not the case.} and defining the
effective action $\gm[\Psi,K_\Phi]$ as the Legendre transform
\cite{Zinn-Justin} of $W[J_\Psi,K_\Phi],$ we find the BRS identity
\begin{equation}
     \idxth \,\bigg(\, 
          {\d\gm\over\d\gm^a_\a}\,{\d\gm\over\d K^{a\a}_\gm } 
          + {\d\gm\over\d C^a}\,{\d\gm\over\d K^a_C}    
          + B^a\,{\d\gm\over\d\hat{C}^a} \,\bigg) = 0 \, .
\label{BRS-id-super}
\end{equation}
To regularize and renormalize the theory, one can think of using
regularization by DReD \cite{Siegel-DReD}. This keeps the advantages
of the superfield formalism. However, there is no a priori reason why
the resulting effective action should satisfy the BRS identity
(\ref{BRS-id-super}), since DReD is not manifestly BRS invariant.
\begin{table}[ht]
\begin{center}
\begin{tabular}{|ll|c|c|} \hline
  \multicolumn{2}{|c|}{external lines} & 1 loop & 2 loops \\ \hline
  $\gm^2$ & &  $\oom=1$ & $\oom=0$   \\ \hline
  $\gm^3$ & $\gm^4$ & $\oom=0$  & \\ \hline
\end{tabular}
\\[9pt]
{\sl Table 3: 1PI superficially divergent super-diagrams for 
$\gm_0.$}
\end{center}
\vspace{-15pt}
\end{table}

\subsection{Component classical action} 

To use ordinary DReG, we turn now to the component formalism. Using
eqs.  (\ref{gauge-comp}), (\ref{ghost-comp}) and (\ref{ext-comp}), it
is not difficult to see that in terms of component fields $S_{\YM},~
S_{\CS},~ S_{\GF}$ and $S_{\ES}$ take the form
\begin{eqnarray}
 S_{\YM} &\igual & {1\over g^2} \idx \, \bigg[
     - {1\over 4}\, F^a_{\m\n} F^{a\m\n}
     - {1\over 2} \, \bar\la^a (\Dslash{} \la)^a\, \bigg]
\label{YM-comp} \\[9pt]
  S_{\CS} &\igual& {1\over g^2} \idx \,\bigg[\,\eps^{\m\n\r}\,
       \Big( \,{1\over 2}\,A^a_\m\partial_\n A^a_\r
            + {1\over 6}\,f^{abc}\, A^a_\m A^b_\n A^c_\r\, \Big)
     - {1\over 2} \> \bar\la^a \la^a\, \bigg]
\label{CS-comp} \\[9pt]
   S_{\GF} &\igual&  \idx \,\bigg\{ \! - b^a \partial_\m V^{a\m} 
       - (\partial^\m\hat{c}^a)\,\,
         \Big( \partial_\m c^a + f^{abc} \,V^b_\m c^c 
       - {i\over 2}\, f^{abc}\,\bar\chi^b\ga_\m\varphi^c \Big) 
\nonumber\\ 
   &\menos& \bar\zeta^a\La^a 
       - \bar{\hat{\varphi}}^a \,\bigg[\, \parslash\varphi^a 
       + f^{abc}\, \Big( \,i\La^b c^c 
       + {i\over 2}\, \ga^\m\chi^b\,\partial_\m c^c 
       + {1\over 2}\,\Vslash^b\varphi^c - {1\over 2}\,H^b\varphi^c 
       + {i\over 2}\,\chi^b\om^c \Big)\, \bigg] 
\nonumber\\
   &\menos& h^a H^a + \hat{\om}^a \, \Big(\,\om^a + f^{abc} \,H^b c^c 
       - {i\over 2}\, f^{abc} \bar\chi^b\varphi^c\,\Big) \bigg\} 
\label{GF-comp}\\[9pt]
   S_{\ES} &\igual& \idx \, \Big[\, 
      i\bar{\ka}^a s\La^a + K^{a\m} sV^a_\m 
      + G^a sH^a + i \bar{\s}^a s\chi^a \ell^a s\om^a
      + i\bar{\t}^a s\varphi^a  + L^a sc^a \,\Big] ~ .
\label{ES-comp}
\end{eqnarray}
Here $F^a_{\m\n}=\partial_\m A^a_\n - \partial_\n A^a_\m + f^{abc}
A^b_\m A^c_\n$ and $D^{ab}_\m=\d^{ab}\partial_\m + f^{acb} A^c_\m$
denote the field strength and the covariant derivative, and $V^a_\m$
and $\La^a$ are given by
\begin{eqnarray}
  & {\ds V^a_\m = A^a_\m
      + {1\over 4}\, f^{abc}\,\bar\chi^b\ga_\m \chi^c } &
\label{V} \\[6pt]
  & {\ds \La^a = \la^a + \parslash\chi^a
      + {1\over 2}\, f^{abc} \,\Aslash^b\chi^c
      - {1\over 2}\, f^{abc}\,H^b \chi^c
      - {1\over 24}\, f^{abc}\,f^{cde}\, \ga^\m\chi^b \,
        (\bar\chi^d\ga_\m\chi^e) ~. } &
\label{Lambda}
\end{eqnarray}
The fields $V^a_\m$ and $\La^a$ have a very simple expression as
superfield projections:
\begin{displaymath}
   V^a_{\a\b} = D_{(\a} \gm^a_{\b)}\,\Big\vert \qquad \qquad
   \La^a_\a =  {i\over 2}\> D_\a D^\b \gm^a_\b \,\Big\vert\>.
\end{displaymath}

The BRS and supersymmetry transformation laws for the components are
obtained from the BRS and supersymmetry transformation laws for the
superfields and the definition of components as projections. After some
algebra, we obtain 
\begin{displaymath}
\begin{array}{rlrl} \qquad
  \gm^a_\a: & s \chi^a =  i\varphi^a - f^{abc} \chi^b c^c   
     & \quad \qquad B^a: &  sb^a   
\\[6pt]         
     & s A^a_\m = (D_\m c)^a && s\zeta^a=0 
\\[3pt]
     & {\ds s H^a = \om^a + f^{abc} H^b c^c 
                  - {i\over 2}\, f^{abc} \bar{\chi}^b\varphi^c} 
     && sh^a=0
\\[6pt]
     & s \la^a = - f^{abc} \la^b c^c && 
\\[6pt]
  \hat{C}^a\!: & s \hat{c}^a = b^a 
  & C^a: & {\ds s c^a = - {1\over 2}\, f^{abc} c^b c^c }
\\[6pt]
  & s\hat{\varphi}^a = i\zeta^a  &
  & s\varphi^a = f^{abc} \varphi^b c^c 
\\[3pt]
  & s\hat{\om}^a = h^a &
  & {\ds s \om^a = - f^{abc} \om^b c^c 
             - {1\over 2}\> f^{abc}\, \bar{\varphi}^b\varphi^c }
\end{array}
\end{displaymath}
for the BRS transformations, and
\begin{displaymath}
\begin{array}{rlrl}
  \gm^a_\a: & \d \chi^a = \Vslash ^a \ee - H^a\ee
  & \quad \qquad B^a: & \d b^a = -\,\bar\zeta^a\ee 
\\[6pt]         
     & \d A^a_\m = \bar\ee \ga_\m \la^a + \bar\ee \,(D_\m \chi)^a  
     & & \d\zeta^a = h^a\ee - \parslash \,b^a\ee 
\\[6pt]
     & \d H^a = -\, \bar\ee \La^a 
     & & \d h^a = \bar\ee\,\parslash\,\zeta^a 
\\[3pt]
     & \ds \d \la^a = -\, {1\over 2}\> \ga^\m\ga^\n F^a_{\m\n}\, \ee
                     + f^{abc} \la^b (\bar\chi^c\ee)   &&
\\[9pt]
  \hat{C}^a\!: & \d \hat{c}^a = i\,\bar{\hat{\varphi}}^a\ee
  & C^a: & \d c^a = i \,\bar\varphi^a\ee 
\\[6pt]
  & \d\hat{\varphi}^a = - i\, \parslash\, \hat{c}^a\ee 
                     + i\, \hat{\om}^a \ee &
  & \d\varphi^a = - i\, \parslash \, c^a\ee + i\,\om^a\ee 
\\[6pt]
  & \d\hat{\om}^a  = i\,\bar\ee\, \parslash\, \hat{\varphi}^a &
  & \d\om^a  = i\,\bar\ee\, \parslash\, \varphi^a 
\\[9pt]
  K_\gm^{a\a}: & \d\ka^a = i\, \Kslash{}^a \ee + i\, \,G^a\ee  
  & K_C^a: & \d \ell^a = \bar\ee\t^a
\\[6pt]
  & \d G^a = i \, \bar\ee\,\parslash\,\ka^a + i\,\bar\ee\s^a &
  & \d\t^a = i \,\parslash \,\ell^a \ee - i\,L^a\ee
\\[6pt]
  & \d K^a_\m = i\,\bar\ee \partial_\m \ka^a + i\,\bar\ee\ga_\m\s^a  &
  & \d L^a = \bar\ee\,\parslash\,\t^a 
\\[3pt]
  & {\ds \d\s^a = \,{i\over 2}\,\ga^\m\ga^\n\ee\,
     (\,\partial_\m K^a_\n - \partial_\n K^a_\m\,)} & &
\end{array}
\end{displaymath}
for the supersymmetry transformations. To understand the gauge
(\ref{gauge-super}) in terms of components, it is convenient to recast
$S_{\GF}$ in eq. (\ref{GF-super}) as
\begin{displaymath}
   S_{\GF} = \idx \> s\, \Big(\! - \hat{c}^a\,\partial_\m V^{a\m}
     + i\, \bar{\hat{\varphi}}^a \La^a - \hat{\om}^a H^a \Big) ~ .
\end{displaymath}
It then becomes clear that $S_{\GF}$ imposes the conditions
\begin{equation}
   \partial_\m V^{a\m}=0 \qquad\qquad \La^a=0 \qquad\qquad H^a=0
\label{gauge-cond}
\end{equation} 
through the Lagrange multipliers $b^a,~\zeta^a,~h^a,$ and that
associated with these conditions there are ghost-antighost pairs
$(c^a, \hat{c}^a),$ $(\varphi^a, \hat{\varphi}^a),$ $(\om^a,
\hat{\om}^a).$ As a check, one may verify that the conditions
(\ref{gauge-cond}) are invariant under the component supersymmetry
transformation laws given above. BRS and supersymmetry invariance for
$S_{\YM}$ and $S_{\CS}$ written in components is straightforward to
check. As regards $S_{\GF}$ and $S_{\ES},$ BRS invariance is trivial
and supersymmetry invariance is easily verified if one uses $[s,\d]=0$
and the supersymmetry transformations for $V^a_\m$ and $\La^a:$
\begin{displaymath}
   \d V^a_\m = \bar{\ee}\ga_\m \La^a 
             - {1\over 2}\>\bar{\ee}\,\Big(\, 
               \ga_\m\ga_\n-\ga_\n\ga_\m\,\Big)
               \, \partial^\n\chi^a
   \qquad \qquad   
   \d \La^a = \partial V^a\ee - \parslash H^a\ee \>.
\end{displaymath}

Introducing real external sources $J^a_\psi$ for the fields $\psi^a =
\chi^a,\, V^a_\m,\, H^a,\, \La^a,\, b^a,\, \zeta^a,\, h^a,\, c^a,\,
\varphi^a,$ $\om^a,\, \hat{c}^a,\, \hat{\varphi}^a, \,\hat{\om}^a,$
denoting by $K^a_\phi$ the external sources for the nonlinear BRS
transforms $s\phi^a~ (\phi^a = \chi^a,\, V^a_\m,\, H^a,\, \La^a,\,
c^a,\, \varphi^a,\, \om^a)$ and following Zinn-Justin
\cite{Zinn-Justin}, it is straightforward to see that the effective
action $\gm\equiv\gm[\psi,K_\phi]$ generating 1PI Green functions of
the fields $\psi^a$ and the sources $K^a_\phi$ satisfies the BRS
identity
\begin{equation}
   \idx \,\bigg(\, 
        \sum_\phi \,{\d\gm\over\d\phi}\,{\d\gm\over\d K_\phi} 
          + b\,{\d\gm\over\d\hat{c}}
          + i\bar\zeta\,{\d\gm\over\d\bar{\hat{\varphi}}}
          + h\,{\d\gm\over\d\hat\om}\,\bigg) = 0 \, .
\label{BRS-id}
\end{equation}
We remark that $\gm$ generates 1PI Green functions for the fields
$V^a_\m$ and $\La^a$ and not for the elementary fields $A^a_\m$ and
$\la^a.$ This is due to the fact that $S_{\ES}$ in eq.
(\ref{ES-comp}) introduces external sources for the BRS variations of
$V^a_\m$ and $\La^a,$ and not for those of $A^a_\m$ and $\la^a.$ To
end up with a BRS identity for an effective action $\gm'$ generating
1PI Green functions for the fields $A^a_\m$ and $\la^a,$ we must
replace $S_{\ES}$ with
\begin{displaymath}
  S'_{\ES} = \idx \, \Big[\, 
      i\bar{\ka}^a s\la^a + K^a sA^a 
    + G^a sH^a + i \bar{\s}^a s\chi^a + \ell^a s\om^a
    + i\bar{\t}^a s\varphi^a  + L^a sc^a \,\Big] ~ .
\end{displaymath}
The problem is then that 
\begin{displaymath}
   \gm'_0={1\over m}\> S_{\YM}+S_{\CS}+S_{\GF}+S'_{\ES}
\end{displaymath}
is not what results from projecting onto components the classical
action $\gm_0$ written in terms of superfields, so we will not be
concerned with the effective action $\gm'$ built upon $\gm'_0.$ Coming
back to the BRS identity (\ref{BRS-id}) and the effective action
$\gm,$ we will need later the explicit form of this identity at one,
two and three loops. To obtain it, we write for $\gm$ a loop expansion
\begin{displaymath}
\gm = \sum_{k=0}^\infty \; \hbar^k \, \gm_k 
\end{displaymath}
and substitute it into eq. (\ref{BRS-id}). This yields
\begin{eqnarray}
  & \Theta \gm_1 = 0 & \label{BRS-1} 
\\[12pt]
  & \Theta \gm_2  + {\ds \idx \,\sum_\phi\, {\d\gm_1\over \d\phi}\, 
         {\d\gm_1\over\d K_\phi} = 0 } & \label{BRS-2}
\\[12pt]
  & \Theta \gm_3 + {\ds \idx\, \sum_\phi\, \bigg( \,
       {\d\gm_1\over \d\phi}\, {\d\gm_2\over\d K_\phi} 
     + {\d\gm_2\over \d\phi}\, {\d\gm_1\over\d K_\phi}\, \bigg) = 0 ~,} 
  & \label{BRS-3}
\end{eqnarray}
where
\begin{equation}
   \Theta = \idx \bigg[ \sum_\phi \, \bigg(\,
            {\d\gm_0\over\d\phi}\,{\d\over\d K_\phi} 
          + {\d\gm_0\over\d K_\phi}\,{\d\over\d\phi}\,\bigg) 
          + b\,{\d\over\d\hat{c}}
          + i\bar\zeta\,{\d\over\d\bar{\hat{\varphi}}}
          + h\,{\d\over\d\hat\om}\,\bigg] 
\label{ST-op}
\end{equation}
is the Slavnov-Taylor operator. The operator $\Theta$ satifies
$\Theta^2\!=\!0$ and $[\Theta,\d]=0,$ a property that we will use in
section 5.

We have two different bases of fields and sources. On the one hand,
there is the basis formed by ${\cal B}_{V\La}=\{V^a_\m,\, \La^a,\,
\chi^a,\ldots, K^{a\m}_V,\,\bar{K}^a_\La,\,\bar{K}^a_\chi,\ldots\},$
and on the other hand, there is the basis formed by ${\cal B}_{A\la}=
\{A^a_\m,\, \la^a,\, \chi^a,\ldots, K^{a\m}_A, \,\bar{K}^a_\la,
\,\bar{K}^a_\chi,\ldots\}.$ We are interested in the effective action
for the fields and sources in the first basis because that is the
effective action which is supersymmetric at tree level. To actually
compute it, we could use Feynman rules for the elements of ${\cal
  B}_{V\La}.$ However, this way to proceed is not convenient, since
the Feynman rules for the Yang-Mills and Chern-Simons actions
$S_{\YM}$ and $S_{\CS}$ are very complicated in terms of $V^a_\m$ and
$\La^a.$ Therefore, we will use the Feynman rules for $A^a_\m$ and
$\la^a,$ and treat $V^a_\m$ and $\La$ as composite fields. Ordinary
power counting for $\gm_0$ shows that the theory is
superrenormalizable, with only the 1PI diagrams in Table 4 being
superficially divergent. We note that there are no superficially
divergent diagrams with external sources as external lines.
\begin{table}[ht]
\begin{center}
\vspace{10pt}
\begin{tabular}{|l l l l l l|c|c|c|} \hline
  \multicolumn{6}{|c|}{external lines}
                     & 1 loop & 2 loops & 3 loops \\ \hline
  $\chi\bar\chi$ &&&&& &  $\oom=2$ & $\oom=1$ & $\oom=0$  \\ \hline
  $\la\bar\chi$  & $A^2$ & $AH$ & $H^2$ &&& ${\ds{\atop \oom=1}}$ &
       ${\ds{\atop \oom=0}}$ & \vspace{-4pt} \\
  $\chi\bar\chi A$ & $\chi\bar\chi H$ & $(\chi\bar\chi)^2$ &&&&&&
       \\ \hline
  $\la\bar\la$ & $\hat{c}c$ & $\hat{\varphi}\bar\varphi$ &
       $\zeta\bar\chi$ &&&&& \\
  $\chi\bar\la A$ & $\chi\bar\la H$ & $A^3$ & $A^2H$ & $AH^2$ &
       $H^3$ & ${\ds{ \atop \oom=0}}$ & & \vspace{-4pt} \\
  $(\chi\bar\chi)\,(\chi\bar\la)$ & $\chi\bar\chi A^2$ &
       $\chi\bar\chi AH$ & $\chi\bar\chi H^2$ &&&&& \\
  $(\chi\bar\chi)^2 A$ & $(\chi\bar\chi)^2 H$ &
       $(\chi\bar\chi)^3$ &&&&&& \\ \hline
\end{tabular}
\\[9pt]
{\sl Table 4: 1PI superficially divergent diagrams for $\gm_0.$}
\end{center}
\vspace{-15pt}
\end{table}
Let us consider a 1PI diagram with $A^a_\m$ and/or $\la^a\!$ external
lines, and let us denote its superficial UV degree of divergence by
$\oom.$ Then it is straightforward to see that the diagrams that
result from replacing one or more of the external $A^a_\m\!$-lines
with $f^{abc}\bar{\chi}^b\ga_\m\chi^c$ and one or more of the external
$\la^a\!$-lines with $f^{abc} \Aslash^b\chi^c,$ $f^{abc} H^b\chi^c$ or
$f^{abc} f^{cde}\ga^\m\chi^b(\bar\chi^d\ga_\m\chi^e)$ all have
superficial UV degree of divergence strictly less than $\oom.$ This is
very simple to see for the fields $A^a_\m$ and $V^a_\m,$ since
$A^a_\m$ couples with a derivative to two other fields $A^a_\m\!$ and
to a ghost-antighost pair $\hat{c}^ac^b$, while
$f^{abc}\bar\chi^b\ga_\m\chi^c$ only couples to two other fields
$\chi^a$ without any derivative. For $\la^a$ and $\La^a,$ replacing
$\la$ by $\parslash\chi^a$ does not introduce worse divergences since
the couplings with one external $\chi^a$ are already taken into
account in Table 4. Replacing $\la^a$ by $f^{abc}\Aslash^b\chi^c,~
f^{abc}H^b\chi^c$ or $f^{abc} f^{cde} \ga^\m\chi^b
(\bar\chi^d\ga_\m\chi^e)$ requires some more analysis, but as can
easily be checked does not lead to worse divergences. All in all,
regarding $V^a_\m$ and $\La^a$ as composite fields does not worsen
power counting.

\section{Dimensional regularization, $\eps^{\m\n\r}$ and BRS
  invariance}

\subsection{Dimensional regularization}

Due to the presence of $\eps_{\m\n\r}$ in the Chern-Simons action,
DReG is not straightforward. To incorporate $\eps_{\m\n\r}$ into the
framework of DReG, we follow ref. \cite{GMR} and use the HVBM
prescription for parity-violating objects, originally introduced by 't
Hooft and Veltman \cite{tHooft} and systematized by Breitenlohner and
Maison \cite{Breitenlohner}. The HVBM prescription defines
$\eps_{\m\r\n}$ in $n\!\geq\!3$ integer dimensions as a completely
antisymmetric object in its indices satisfying the relations
\cite{Breitenlohner}
\begin{equation}
\eps_{\m_1\m_2\m_3} \eps_{\n_1\n_2\n_3} = \sum_P\; (-1)^{|P|}\,
    \tg_{\m_1\n_{P1}}\, \tg_{\m_2\n_{P2}}\, \tg_{\m_3 \n_{P3}} 
\qquad\qquad 
\eps_{\m_1\m_2\m_3}\hg^{\m_3\m_4} = 0 \, ,
\label{n-eps}
\end{equation}
where all indices run from 0 to $n\!-\!1,$ the sum is extended over
all permutations $(1,2,3) \to (P1,P2,P3),$ $|P|$ is the order of the
permutation $P,$ $g_{\m\n}$ is the metric on $\RR^n,$ and $\tg_{\m\n}$
and $\hg_{\m\n}$ are its projections onto $\RR^3$ and $\RR^{n-3}.$ In
other words, $\eps^{\m\n\r}$ is treated as a three-dimensional object.
In what follows, we will regard objects with tildes and hats as
projections onto $\RR^3$ and $\RR^{n-3},$ respectively. That is,
$\tp_\m=\tg_{\m\n}p^\n, ~\hp^2=\hg_{\m\n}p^\m p^\n,~ etc.$ Hatted
objects vanish at $n=3$ and are usually called evanescent in the
literature. We remark that the HVBM definition of $\eps^{\m\n\r}$
outside three dimensions is the only algebraically consistent one
known to date \cite{Bonneau}.  

Once we have an algebraically consistent definition for
$\eps^{\m\n\r}$ in $n$ dimensions, we dimensionally regularize the
theory as follows \cite{GMR}:
\begin{itemize}
\item[(i)] First, we extend the Feynman rules from three dimensions to
  $n$ dimensions.
\item[(ii)] Next we construct $n\!$-dimensional 1PI diagrams and use
  the techniques of refs. \cite{tHooft} and \cite{Collins} to continue
  $n$ to complex values $d.$ This replaces every three-dimensional
  1PI diagram in the original theory with a dimensionally regularized
  diagram defined in terms of dimensionally regularized integrals.
  It must be emphasized that, when continuing $n$ from integer to
  complex values, the quantities $p^2,~ p^\m, g_{\m\n}$ and
  $\eps^{\m\n\r}$ cease to have meaning as scalars, vectors and
  tensors and are defined only through their algebraic relations
  \cite{Breitenlohner}.
\item[(iii)] Finally, we compute the dimensionally regularized
  integrals entering in a dimensionally regularized Feynman diagram and
  analytically continue the result to $d=3.$ This defines the value of
  the dimensionally regularized 1PI diagrams. As usual, computation of
  dimensionally regularized integrals entails a Wick rotation to
  euclidean momentum space.
\end{itemize}
\noindent
The extension of the Feynman rules from three to $n$ dimensions is
obtained as in QCD, except the propagator of the gauge field, which
deserves some attention. In three dimensions, the quadratic part of
the action in $A^a_\m$ and $b^a$ has in momentum space the form
\begin{displaymath}
  -\,{1\over 2} \idp \Big[\, A^a_\r(p) \, K^{\r\m}(p) \, A^a_\m(-p) 
              +\, b^a(p)\,p^\r\,A^a_\r(-p)\,\Big] ~,
\end{displaymath}
where 
\begin{displaymath}
   K^{\r\m}(p) = {1\over g^2}\> \bigg[ - \eps^{\r\s\m}\,p_\s 
               + {i\over m}\; \big(\, p^2 g^{\r\m} - p^\r p^\m \,
               \big)\, \bigg] ~ .
\end{displaymath}
This defines the kinetic matrix of $A^a_\m$ and $b^a$ as
\begin{displaymath}
   T(p) \; = \;  \pmatrix{ ~ K^{\r\m}(p) ~ & ~ - p^\r \cr
               ~ p^\m ~ & 0 \cr} ~ .
\end{displaymath}
Inverting $T(p)$ in three dimensions, we obtain the three-dimensional
propagators for the fields $A^a_\m$ and $b^a,$ namely
\begin{equation}
   {\langle\,A^a_\m(p)\, A^b_\n(-p)\,\rangle}^{(3)}_0 
                = \d^{ab}\,D_{\m\n}(p) 
\label{3-prop}
\end{equation}
and
\begin{displaymath}
   {\langle\,A^a_\m(p)\,b^b(-p)\,\rangle}^{(3)}_0
                = \d^{ab}\; {p_\m\over p^2} ~,
\end{displaymath}
with $D_{\m\n}(p)$ given by
\begin{equation}
  D_{\m\n}(p) = - \, 
       { g^2\,m \over p^2 \,(p^2\!+\!m^2\!-\!io)}~
       \Big(\, m\, \eps_{\m\r\n}\, p^\r + i\,p^2 g_{\m\n} 
           - i\, p_\m p_\n \,\Big) ~ .
\label{Delta}
\end{equation}
According to the arguments in ref. \cite{Breitenlohner}, for BRS
invariance to be manifestly preserved, the $n\!$-dimensional
propagators for $A^a_\m$ and $b^a$ should be computed by inverting the
kinetic matrix $T(p)$ in $n$ dimensions. Doing this and using for
$\eps^{\m\n\r}$ the HVBM definition given above, we obtain
\begin{displaymath}
   {\langle\,A^a_\m(p)\, A^b_\n(-p)\,\rangle}^{(n)}_0 
                = \d^{ab}\,\De_{\m\n}(p) \qquad\qquad
   {\langle\,A^a_\m(p)\,b^b(-p)\,\rangle}^{(n)}_0
                = \d^{ab}\; {p_\m\over p^2}~ ,
\end{displaymath}
where $\De_{\m\n}(p)$ has the form
\begin{equation}
\begin{array}{l} {\ds
  \De_{\m\n}(p) = - \,  
    {g^2\,m\over (p^2\!-\!io)^2 + m^2\tp^2}~\bigg[ \, 
        m\,\eps_{\m\r\n}\,p^\r + i\, p^2 g_{\m\n} - i\,p_\m p_\n }
\\[12pt] \phantom{\ds D_{\m\n}(p) = \; } {\ds 
      + \, { i\,m^2\over p^2\!-\!io}~ \bigg( \tp^2 \hg_{\m\n} 
         + {\hp^2\over p^2}\, p_\m p_\n - p_\m \hp_\n - \hp_\m p_\n 
         + \hp_\m \hp_\n \bigg)\,\bigg]  ~ . }
\end{array}
\label{D-prop}
\end{equation}
The complicated dependence of $\De_{\m\n}(p)$ on $\tp^\m$ and $\hp^\m$
arises from the fact that $\eps^{\m\n\r}$ in $n$ dimensions transforms
covariantly under $so(1,2) \times  so(n-3)$ rather than under the
full Lorentz group $so(1,n),$ and is the price to pay for manifest BRS
invariance.  We emphasize that we want manifest BRS invariance, since
our proof of perturbative finiteness in the next section is based on
the fact that DReG manifestly preserves BRS invariance.

To avoid a propagator as involved as $\De_{\m\n}(p),$ one might wish
to simply take the expression for $D_{\m\n}(p)$ in eq. (\ref{Delta})
and regard $\eps_{\m\n\r}$ as defined above and $p^\m$ and $g_{\m\n}$
as $\!n\!$-dimensional. This way to proceed simplifies the
calculations but does not manifestly preserve BRS invariance. To see
this, let us consider the matrix propagator corresponding to the
propagator $D_{\m\n}(p),$ {\it i.e.}
\begin{displaymath}
   M_D(p)\; =\; 
   \pmatrix{ ~ D_{\m\n}(p) ~ & ~ {\ds {p_\m\over p^2}} ~ \cr
             ~ - {\ds {p_\n\over p^2}} ~ & ~ 0 ~}   
\end{displaymath} 
and let us invert it in $n$ dimensions. The result is not the kinetic
matrix $T(p)$ but rather
\begin{displaymath}
   T_D (p) \; = \; 
     \pmatrix{ ~ K^{\r\m}(p)+B^{\r\m}(p) ~ & ~ -p^\r \cr
               ~ p^\m ~ & 0 \cr} ~,
\end{displaymath}
where
\begin{displaymath}
\begin{array}{l}
   {\ds  B^{\r\m}(p) = -\, {1\over g^2} ~ 
        {m\over (p^2\!-\!io)^2 + m^2\tp^2} ~ \bigg[ \; 
           \hp^2\, \Big(\,m\,\eps^{\r\s\m}p_\s
         - i\, p^2 g^{\r\m} + i\, p^\r p^\m\,\Big) }
\\[12pt] \phantom{B^{\r\m}(p) = \,} 
   {\ds -\, i\>(p^2+m^2)\;\Big(\,{\hp^2\over p^2}\, p^\r p^\m
     + \tp^2 \hg^{\r\m}  - p^\r \hp^\m - \hp^\r p^\m + \hp^\r \hp^\m
     \Big)\, \bigg] ~. }
\end{array}
\end{displaymath}
The arguments in ref. \cite{Breitenlohner} then imply that the BRS
identity for a dimensionally regularized 1PI Green function
$G(p_e)\equiv\!  \!G(p_1,\ldots,p_E)$ computed with the propagator
$D_{\m\n}(p)$ contains an extra BRS-violating term which arises
because the evanescent contribution
\begin{equation}
    {i \over 2} \int \! d^n\!x \, d^n\!y \, 
         A^a_\m(x)\, B^{\m\n}(x-y) \, A^a_\n(y) 
\label{evanescent}
\end{equation}
to the action is not BRS invariant. Indeed, the BRS variation of
(\ref{evanescent}) produces an evanescent vertex $O^{abc}_{\m\n}(p,q)
= f^{abc}\, \big[\,B_{\m\n}(p) - B_{\n\m}(q)\,\big]$ in the
dimensionally regularized perturbation series for $G(p_e).$ Formally,
the breaking can be written as
\begin{equation}
  \De G(p_e) = \lim_{d\to 3} \; [\, G(p_e) \int \! d^d\!p \, d^d\!q\;
  f^{abc}\, A^a_\m(-p)\, B^{\m\n}(p) \, A^b_\n(q)\, c^c(p-q) \,]_D ~,
\label{breaking}
\end{equation}
where $[\cdots]$ denotes 1PI and the subscript $D$ refers to the
propagator $D_{\m\n}(p).$ Pictorially, the vertex
$O^{abc}_{\m\n}(p,q)$ is depicted in Fig. 1a and the breaking in Fig
1b.  Note that the difference between $\De_{\m\n}(p)$ and
$D_{\m\n}(p)$ is again a purely evanescent object, since 
\begin{equation}
   \De_{\m\n}(p) = D_{\m\n}(p) + R_{\m\n}(p) 
\label{Delta-R}
\end{equation}
and
\begin{displaymath}
\begin{array}{l} {\ds
   R_{\m\n}(p) = -\,
     {g^2 \,m^3 \over (p^2\!-\!io)^2 + m^2\,\tp^2}~
       \bigg[ \, {\hp^2\over p^2 \, (p^2\!+\!m^2\!-\!io)} ~ \Big(
          m\,\eps_{\m\r\n}\,p^\r + i\, p^2 g_{\m\n}
        + {i\,m^2\over p^2\!-\!io} \,p_\m p_\n\Big) }
\\[12pt] \phantom{\ds R_{\m\n}(p) =
      -\, { g^2\,m^3 \over (p^2\!-\!io)^2 + m^2\,\tp^2}\, }
     {\ds +\, {i \over p^2\!-\!io} ~ \Big(\, \tp^2 \hg_{\m\n}
          - p_\m \hp_\n - \hp_\m p_\n  + \hp_\m \hp_\n \Big)\,
          \bigg] ~  }
\end{array}
\end{displaymath}
vanishes at $n\!=\!3.$ The identity (\ref{Delta-R}) will be used in
the next section. It is very important to note that $\De_{\m\n}(p)$
and $D_{\m\n}(p)$ have both UV degree -2, whereas $R_{\m\n}(p)$ has UV
degree -4. As concerns IR power counting, $\De_{\m\n}(p)$ and
$R_{\m\n}(p)$ have IR degree -2, and $D_{\m\n}(p)$ has IR degree -1.
\begin{center}
\epsfig{file=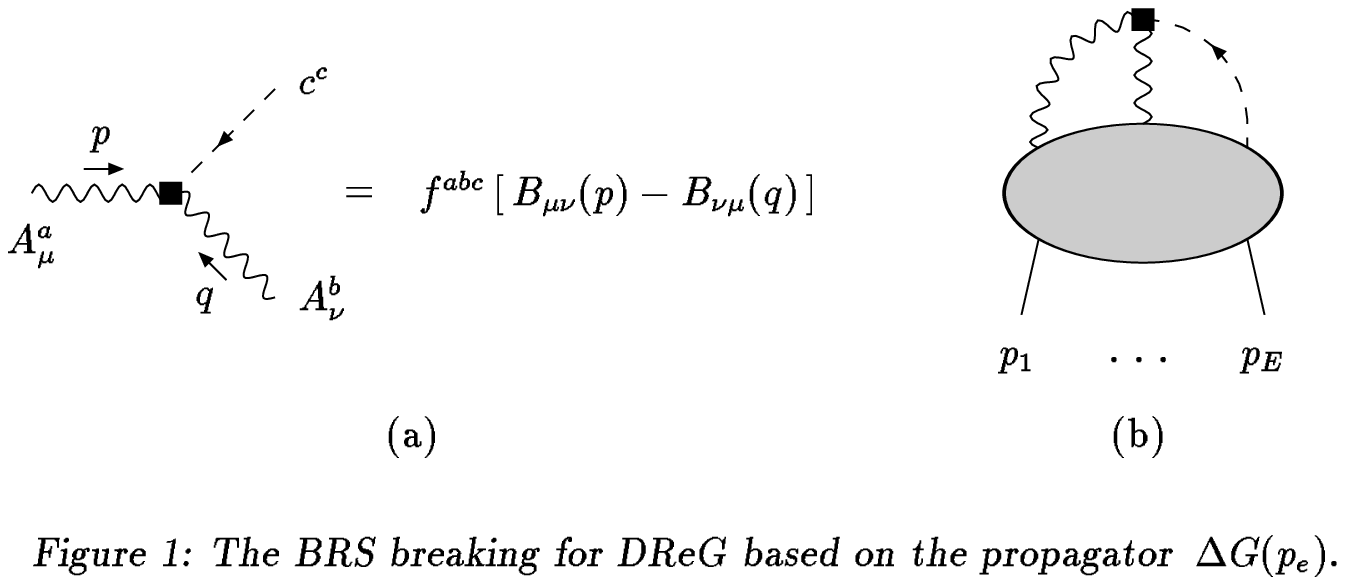} 
\end{center}

\subsection{Regularization by dimensional reduction}

Let us very briefly recall the basics of regularization by DReD. In
the original formulation of DReD \cite{Siegel-DReD}, all the fields
and matrices are kept three-dimensional, so that the Dirac algebra of
the Feynman diagrams is performed in three dimensions. The momenta
however are continued in the sense of ordinary DReG to $d\!<\!3.$ This
way to proceed manifestly preserves supersymmetry since the Dirac
algebra is performed in three dimensions and thus Fierz identities
remain valid. Unfortunately, the propagator for the gauge field that
results from this prescription, namely $D_{\m\n}(p)$ in eq.
(\ref{3-prop}) with $p^\m\!\!$ $d\!$-dimensional and $\eps_{\m\n\r}$
and $g_{\m\n}$ three-dimensional does not formally admit an inverse in
$d\!<\! 3$ dimensions. This implies that DReD does not manifestly
preserve BRS invariance.

As is well known, DReD is algebraically inconsistent because different
contractions of three or more $\eps^{\m\n\r}$ factors yield different
results in $d\!<\! 3$ dimensions \cite{Siegel-incon}.  However, in our
model, this inconsistency is absent since the contributions with
three or more factors $\eps^{\m\n\r}$ are finite by power counting,
due to the fact that for large momenta the Yang-Mills action gives
the dominant contribution.

\section{Perturbative finiteness}

In what follows we prove that all dimensionally regularized 1PI Green
functions of the fields $\psi^a$ and the sources $K^a_\phi$ are finite
to all orders in perturbation theory, meaning that no poles arise in
them when the regulator $d$ is taken to 3. Before presenting the
proof, let us recall the following property of dimensionally
regularized integrals, due to Speer \cite{Speer}. Consider the
dimensionally regularized integral
\begin{equation} 
  I_{\m_1\ldots\m_N}(p_e,m,d) = \iddq ~ { q_{\m_1}\cdots q_{\m_N}
     \over \prod_{r,s} \, (Q_r^2)^{n_r} \,(Q_s^2\!+\!m^2)^{n_s} } ~ ,
\label{Speer}
\end{equation}
where $Q^\m_r$ and $Q^\m_s$ are linear combinations of the loop
momentum $q^\m$ and the external momenta $p_e^\m,$ and $n_r$ and $n_s$
are nonnegative integers. Then analytic continuation of
$I_{\m_1\ldots\m_N}(p_e,m,d)$ to $d\to n_0,$ with $n_0$ odd, does not
produce poles, even though $I_{\m_1\ldots\m_N}(p_e,m,d)$ might not be
finite by power counting at $d\!=\!n_0.$ We call dimensionally
regularized integrals of this type Speer integrals. It is important to
note that the denominator in the integrand in eq.  (\ref{Speer}) is
Lorentz covariant. Hence, dimensionally regularized integrals with
factors in the denominator of the type $p^4\!+\!m^2 \tp^2$ arising
from internal gauge lines are not of Speer type. Let us now proceed
with the proof.

\subsection{One loop}

We recall from section 2 that superficially divergent one-loop 1PI
diagrams for component fields have $\oom=0,1,2.$ Furthermore, some
very simple power counting show that all 1PI one-loop diagrams have IR
degree $\uom\geq 1.$ Let us consider a superficially divergent
one-loop 1PI diagram and call $\cD(d)$ to the corresponding
dimensionally regularized diagram.  If the diagram does not have
internal gauge lines, $\cD(d)$ is made of dimensionally regularized
integrals of Speer type and hence does not give rise to poles as $d\to
3.$ So we only have to consider superficially divergent diagrams with
internal gauge lines. We distinguish two cases: $\oom=2$ and
$\oom=0,1.$

{\it Case $\oom=2.$} The only one-loop 1PI diagrams with $\oom=2$ are
the $\chi^a\bar\chi^b$ selfenergy graphs. It is easy to see from
eq. (\ref{GF-comp}) that there are no such graphs with internal gauge
lines.

{\it Case} $\!\oom=0,1.$ Using for the propagator of each gauge line
the decomposition in eq. (\ref{Delta-R}), we write $\cD(d)$ as the sum
of two contributions: $\cD(d) = \cD_D(d) + \cD_R(d).$ The first one,
$\cD_D(d),$ arises from replacing every propagator $\De_{\m\n}(p)$
with $D_{\m\n}(p)$ and is of Speer type. The second one, $\cD_R(d),$
contains contributions with one or more $R_{\m\n}$ and is not of Speer
type. Since the original diagram had $\oom\leq 1$ and $\uom\geq 1,$
and every $R_{\m\n}$ decreases $\oom$ by two units and leaves $\uom$
unchanged, $\cD_R$ is made of dimensionally regularized integrals
which are finite at $d=3$ and which are at least linear in
$\hg_{\m\n}:$
\begin{equation}
   \hg_{\m_1\n_1}\ldots\hg_{\m_N\n_N} \int\! d^d\!q ~
     { q^{\n_1}\ldots q^{\n_N} \over \prod_{r,s,t} (Q_r^2)^{n_r} \,
       (Q_s^2\!+\!m^2)^{n_s}\, (Q_t^4\!+\!m^2 \tilde{Q_t}^2)^{n_t} }
   \qquad\qquad N\geq 1.
\label{type}
\end{equation}
Integrals of this type vanish as $d\to 3$ \cite{Collins}. Thus
$\cD_R(d)\to 0$ as $d\to 3$ and in this limit we are left only with
the Speer-type contribution $\cD_D(d),$ which does not generate
poles.

To prove one-loop finiteness, it would have been enough to consider
DReD instead of DReG and use that all one-loop dimensionally
regularized integrals arising from DReD are of Speer type, hence free
of poles as $d\to 3.$ This would have avoided the discussion on
evanescent contributions. However, DReD and supersymmetry do not by
themselves imply finiteness at higher loops, whereas DReG and BRS
invariance do (see below).

Note that we have not only proved one-loop finiteness but also that to
compute the limit of physical interest $d\to 3$ we can replace the
propagator $\De_{\m\n}(p)$ with the propagator $D_{\m\n}(p).$
Furthermore, since $D_{\m\n}(p)$ generates Speer integrals and these
do not give rise to poles as $d\to 3,$ we can equally well perform the
Lorentz algebra of the Feynman diagrams directly in three dimensions.
So, all in all, DReG and DReD give the same one-loop Green functions.

\subsection{Higher loops}

At two loops we proceed differently since Speer's result only holds at
one loop. Let us assume that there are divergences at two loops when
$d\to 3.$ Then the two-loop correction $\gm^{\rm DReG}_2$ to the
effective action will consist in the limit $d\to 3$ of a divergent
part $\gm^{\rm DReG}_{\rm 2,div}$ and a finite part $\gm^{\rm
  DReG}_{\rm 2,fin}.$ Since $\gm^{\rm DReG}_2$ satisfies the BRS
identity (\ref{BRS-2}) and $\gm^{\rm DReG}_1$ is finite, the divergent
part $\gm^{\rm DReG}_{\rm 2,div}$ satisfies the equation $\Theta
\gm^{\rm DReG}_{\rm 2,div}=0.$ Because 1PI Feynman diagrams involving
external sources are finite by power counting and there are no
one-loop subdivergences, $\gm^{\rm DReG}_{\rm 2,div}$ does not depend
on the external sources and $\Theta \gm^{\rm DReG}_{\rm 2, div}=0$
reduces to $s\gm^{\rm DReG}_{\rm 2, div} =0,$ with $s$ the BRS
operator. We recall from Table 4 that all two-loop superficially
divergent 1PI diagrams with $A^a_\m$ and $\la^a\!$ external lines have
$\oom=0.$ Hence the most general form of $\gm^{\rm DReG}_{\rm 2,div}$
compatible with power counting is
\begin{displaymath}
  \gm^{\rm DReG}_{\rm 2,div}=\,{1\over d-3} ~ P_{\oom_2} ~,
\end{displaymath}
where $P_{\oom_2}$ is given by 
\begin{equation}
\begin{array}{rcl}
    P_{\oom_2} &\igual& m \idx \,\Big[\, \a_1\,m\, \bar{\chi}^a \chi^a   
        + \a_2 \,\bar{\chi}^a \parslash \chi^a
        + \a_3 \,\bar{\chi^a} \la^a + \a_4\, A^a A^a + \a_5\, H^a H^a 
\\[9pt]
    &\mas& \a_6\, f^{abc} \bar{\chi}^a\Aslash^b\chi^c
        + \a_7\, f^{abc} f^{cde} (\bar\chi^a\ga^\m\chi^b)\,
                  (\bar\chi^d\ga_\m\chi^e)\,\Big]  \label{Y}
\end{array}
\label{P-2}
\end{equation}
and $\a_1,\ldots,\a_7$ are numerical coefficients. In writing the
expression for $\gm^{\rm DReG}_{\rm 2,div}$ we have used that two-loop
contributions to 1PI Green functions arising from evanescent operators
$R_{\m\n}(p)$ are finite by power counting and therefore free of
poles.  The terms in $P_{\oom_2}$ correspond to all Lorentz invariant
two-loop divergences that can be constructed from Table 4 with
$\oom_2$ derivatives. The equation $s\gm^{\rm DReG}_{\rm 2,div}=0$ is
an equation in the coefficients $\a_i,$ whose only solution is
$\a_i=0.$ This implies $\gm^{\rm DReG}_{\rm 2,div}=0$ and proves
finiteness at two loops.

The proof at three loops is analogous. In this case, the would-be
three-loop divergent contribution $\gm^{\rm DReG}_{\rm 3,div}$ to the
effective action has the form 
\begin{displaymath}
  \gm^{\rm DReG}_{\rm 3,div} = {\a\,m^2\over d-3} 
     \idx \bar{\chi}^a \chi^a
\end{displaymath}
and satisfies the equation $\Theta \gm_{\rm 3,div}^{\rm DReG}=0,$
whose only solution is $\gm^{\rm DReG}_{\rm 3,div}=0.$ At higher loops
finiteness is trivial, since there are no subdivergences and all 1PI
are superficially convergent. Since all 1PI Green functions for the
elementary fields are finite to all orders in perturbation theory, we
conclude that the beta functions of $g$ and $m$ and the anomalous
dimensions of the elementary fields vanish to all orders in
perturbation theory.

Let us finally see why DReD and supersymmetry do not imply finiteness
at two, hence at higher loops. Suppose we use DReD, instead of DReG.
Since it preserves supersymmetry, the divergent part $\gm^{\rm
  DReD}_{\rm 2,div}$ of the resulting two-loop effective effective
action must be supersymmetric. In other words, it should have the form
$\gm^{\rm DReD}_{\rm 2,div}= {1\over d-3} P^{\rm susy}_{\oom_2},$ with
$ P^{\rm susy}_{\oom_2}$ a supersymmetry invariant. From $P_{\oom_2}$
in eq. (\ref{P-2}) above one can construct the supersymmetry invariant
\begin{equation}
   P_{\oom_2}^{\rm susy} = \a\> m \idx \, \Big[\, 
         {1\over 2}\> \bar{\chi}^a \parslash \chi^a 
       + \bar{\chi^a} \la^a  + A^a A^a - H^a H^a
       - {1\over 48} \> f^{abc} f^{cde} (\bar\chi^a\ga^\m\chi^b)\,
                  (\bar\chi^d\ga_\m\chi^e)\,\Big] ~ ,
\label{susy-invariant}
\end{equation}
with $\a$ an arbitrary numerical coefficient. Hence, supersymmetry by
itself does not prove finiteness. This is why we have used DReG and BRS
invariance. 

\section{The effective action} 

Since the theory is finite, every regularization method defines a
renormalization scheme. Let us consider the following two {\it
renormalization schemes}: scheme $\cR^{\rm DReG}$ uses as regulator
DReG and performs no subtractions, and scheme $\cR^{\rm DReD}$ uses
DReD and performs no subtractions. We want to prove that the
difference $\De\gm= \gm^{\rm DReG} - \gm^{\rm DReD}$ between the
corresponding effective actions is zero. We have already seen in
section 4 that this is indeed the case at one loop. So let us consider
the two-loop case. 

According to general results from renormalization theory \cite{Hepp}
\cite{Breitenlohner}, the difference $\De\gm_2$ at two loops can at
most have the form (\ref{diff-2}), with $P_{\oom_2}$ as in eq.
(\ref{P-2}).  Since DReG preserves BRS invariance manifestly,
$\gm^{\rm DReG}_2$ satisfies the BRS identity at two loops
(\ref{BRS-2}). Substituting eq. (\ref{diff-2}) in eq.  (\ref{BRS-2}),
acting from the left with the supersymmetry generator $\d$ and using
$\gm^{\rm DReG}_1=\gm^{\rm DReD}_1$ and $[\Theta,\d]=0,$ we obtain
$\Theta\d P_{\oom_2}=0.$ Since $P_{\oom_2}$ does not depend on
external sources and $\d$ acting on the components of the gauge
multiplet does not produce external sources, $\d P_{\oom_2}$ is
independent of external sources. Therefore the equation $\Theta\d
P_{\oom_2}=0$ reduces to $s\d P_{\oom_2}=0,$ which is an equation for
the coefficients $\a_i,$ $s$ being the BRS operator. Since $\d
P_{\oom_2}$ depends polynomially on the components of the gauge
multiplet and their derivatives and has an overall factor of $m,$ any
nontrivial $\d P_{\oom_2}$ satisfying $s\d P_{\oom_2}=0$ should be $m$
times a BRS invariant of mass dimension two. However, there are no
such invariants.  Hence, $\d P_{\oom_2}=0.$ The only supersymmetry
invariant that can be formed from $P_{\oom_2}$ is $P_{\oom_2}^{\rm
  susy},$ which in terms of superfields takes the form in eq.
(\ref{susy-invariant-super}). At this point we have exhausted all
information given by BRS symmetry and supersymmetry. The only way left
to determine the value of the coefficient $\a$ in $P_{\oom_2}$ is to
compute it using Feynman diagrams. We do this below and find that
$\a\!=\!0.$

At three loops, the difference $\Delta \gm_3$ is 
\begin{displaymath}
  \gm^{\rm DReG}_{\rm 3,div} = {\a \,m^2} 
     \idx \bar{\chi}^a \chi^a ~.
\end{displaymath}
Since $\Delta\gm_3$ is not BRS invariant, nor supersymmetric, the same
arguments as used at the two-loop level are now powerful enough to
conclude that $\a\!=\!0$ without the need of any explicit computation.
At higher loops, the difference $\De\gm$ vanishes since at one, two
and three loops it vanishes and there are no overall divergences by
power counting.

We are left with the computation of the coefficient $\a.$ To calculate
it, it is enough to evaluate the difference between the contributions
from DReG and DReD to one of the five proper functions in eq.
(\ref{susy-invariant}). The simplest case to compute is the selfenergy
of the field $H^a.$ The vertices with an $H$ are
\begin{displaymath}
   H\zeta\chi \qquad H\hat{\varphi}\varphi \qquad
   H\hat{\om} c \qquad H\hat{\varphi}\chi c \>.
\end{displaymath}
Using these vertices, one can construct two-loop 1PI diagrams with the
six topologies in Fig. 2. In fact, no graphs with the topology of Fig.
2a can be constructed, since there is no four-point vertex containing
the fields $H,\,\varphi$ and $\hat{c}$ (note that $\hat{\varphi}$ only
propagates in $\varphi$ and $c$ into $\hat{c}.)$ The topologies in
Figs. 2b and 2c, being products of one-loop topologies, give the same
contributions in DReG as in DReD, hence they do not contribute to
$\a.$ We are thus left with the topologies in Figs.  2d, 2e and 2f.
\begin{center}
\epsfig{file=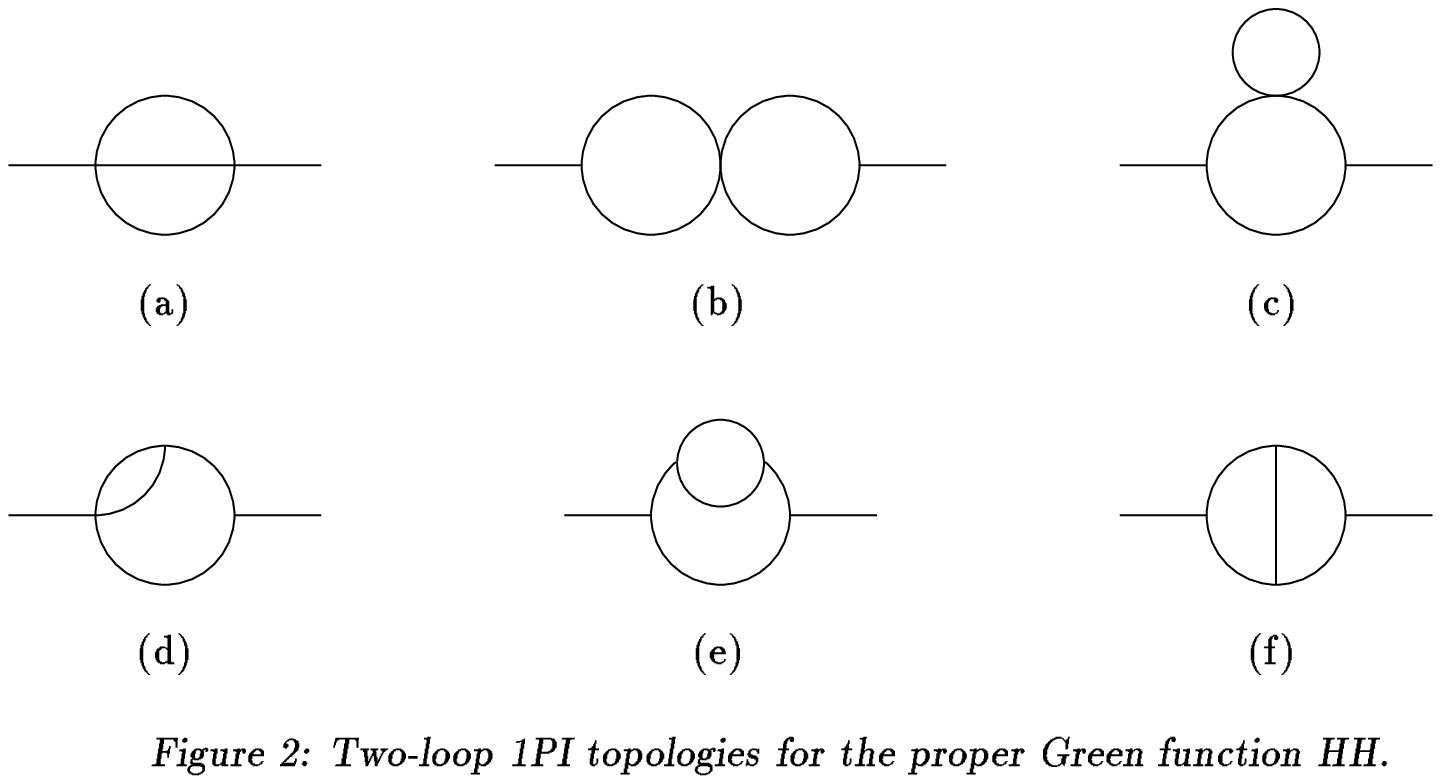}
\end{center}

Since one-loop subdiagrams give the same contributions in DReG as in
DReD, we only need to consider the overall divergent parts of the
two-loop diagrams. Since the diagrams are only logarithmically
divergent by power counting, we may set the external momentum $p^\m$
and the mass $m$ equal to zero in the numerators (except, of course,
of the overall factor $m).$ Because every epsilon in the propagator
and in the three-vertex of the gauge field $A^a_\m$ reduces the
overall degree of divergence by one unit, the overall divergent part
of every diagram is epsilon-independent.  All this gives for the
overall divergent part of every diagram an expression of the form
\begin{displaymath}
   \int \!{d^dk\over (2\pi)^d} ~ {d^dq\over (2\pi)^d}~ 
      { N(k,q) \over D(k,q,p,m)} ~.
\end{displaymath}
The numerator $N(k,q)$ always contains a trace over a fermion loop.
This is obvious for those diagrams in which $H$ couples to fermions.
The only vertex where $H$ does not couple to fermions is the vertex
$H\hat{\om}c,$ but in this case $\hat{\om}$ propagates into $\om$ and
now $\om$ couples to fermions. In fact, no two-loop diagram with this
structure can be constructed. As far as the diagrams with internal
gauge lines are concerned, they only occur in topology 4e and closer
inspection reveals that their contributions separately cancel.
Anyhow, even if they had not cancelled, one could have decomposed the
propagator $\De_{\m\n}(p)$ into a covariant part $D_{\m\n}(p)$ and an
evanescent part $R_{\m\n}(p).$ The latter part yields an evanescent
contribution which is finite by power counting and hence vanishes.
Thus the contributions of both DReG and DReD are the same except for
the trace over the fermions. The trace of a sum of products of
$\qslash$ and $\kslash$ can always be written as $d\!$-dimensional
scalar products $k^2,\,kq$ and $q^2$ times an overall trace of the
unit matrix. Since this trace is different in DReG and DReD, after
summing over diagrams, $\a$ can be written as
\begin{displaymath}
   \a = \Big( {\rm tr}_{\ss\rm \,DReG} \, \unit 
            - {\rm tr}_{\ss\rm \,DReD} \,\unit\Big) 
      \int \!{d^dk\over (2\pi)^d} ~ {d^dq\over (2\pi)^d}~ 
      { f(k^2,kq,q^2) \over D_T(k,q,p,m)} ~,
\end{displaymath}
where $f(k^2,kq,q^2)$ is a polynomial of its arguments. Since we have
already shown that the theory is finite, the integral is finite and
therefore the difference due to the trace vanishes in the limit $d\to
3.$ Hence $\a=0.$

\section{Further comments}

We conclude with a few comments.

1. The equality of the two effective actions considered in this paper
is not explained by the standard theorems of renormalizable quantum
field theory. One possible explanation might be that there exists a
third, as yet unknown, symmetry of the model. Another explanation
might be that the existing theorems of local quantum field theory can
be sharpened for finite models which are superrenormalizable by power
counting and which have symmetries.

2. For the purely bosonic theory, it has been claimed without proof
that since the theory is superrenormalizable, subtleties due to the
epsilon tensor should not matter \cite{Pisarski}. This can easily be
proved for both the bosonic and the supersymmetric theories. Clearly,
such subtleties may only arise from superficially divergent graphs. At
one loop, since there are no 1PI graphs with $\oom\!=\!2$ containing
internal gauge lines nor three-gauge vertices, and since every epsilon
occuring in a 1PI graph decreases the UV superficial degree of
divergence of the graph by one unit, only 1PI graphs with $\oom\!=\!1$
may produce epsilon ambiguities. The ambiguities, if any, will be
linear in $\eps^{\m\n\r}$ and independent of the external momenta,
since they must depend polynomially on the external momenta and arise
from Feynman integrals that are logarithmically divergent. It is very
easy to see that, out of just one $\eps^{\m\n\r}$and nothing else, no
Lorentz invariant can be constructed for the 1PI Green functions in
Table 4 that have $\oom=1$ at one loop. Hence, there are no epsilon
ambiguities at one loop. The same arguments show that this is also the
case at two and higher loops. The source of different results for
different regularization methods is not actually the epsilon, but the
parity-even sector of the theory.

3. In this article we have considered DReG and DReD, but one can also
consider a covariantized DReG method based on the naively
covariantized $n\!$-dimensional propagator $D_{\m\n}(p)$ in eq.
(\ref{Delta}). For the purely bosonic theory it has been shown that
this `covariantized' DReG gives the same effective action as DReG
\cite{GMR}. A straightforward generalization of the arguments given
there shows that this is also the case for the superysmmetric case we
have considered here.

4. Our analysis relies on the fact that our three-dimensional model is
superrenormalizable by power counting and finite. There exist several
one-loop finite supersymmetric models in four dimensions, and
$N\!=\!4$ Yang-Mills theory is even finite to all loop orders. It
would be interesting to apply the methods developed in this paper to
these models. A hint that also for these models under certain
conditions DReG and DReD could give the same results is provided by
the one-loop analysis of $N\!=\! 1$ supersymmetric Yang-Mills theory
in four dimensions in ref. \cite{CJvN}. Note, however, that in this
reference a nonsupersymmetric gauge was used. 

5. In addition to Yang-Mills-Chern-Simons models, there exist
Einstein-Chern-Simons models (topologically massive gravity)
\cite{Deser} \cite{Lerda}. Perhaps our methods can be applied in these
cases \cite{Kleppe}.

6. Our analysis used component graphs and not supergraphs because we
needed DReG to prove finiteness and DReG cannot be formulated for
superfields. In fact, since the classical Yang-Mills and Chern-Simons
actions contain many terms when written in terms of the spinor
connection $\gm^a_a$ [see eqs.  (\ref{YM-super}, (\ref{CS-super})],
using supergraphs is not that advantageous. Also note that there are
no nonrenormalization theorems in three dimensions because there are
no chiral superfields.

7. Chern-Simons theory by itself has the problem that there exists no
propagator in $n\!\geq\!3$ dimensions for the gauge field, even for a
nonvanishing gauge-fixing parameter \cite{Martin}. Hence, for this
model, DReG cannot be formulated in a manifestly BRS invariant way.
One way to overcome this is to add to the Chern-Simons action the
Yang-Mills term \cite{GMR} \cite{Martin} or any other gauge-invariant
parity-even term \cite{GMR2}. Conversely, starting with Yang-Mills
theory, one encounters IR divergences on shell.  A way to regularize
these divergences is to add a Chern-Simons term to the Yang-Mills
action \cite{Schonfeld}. Hence in both cases we end up with
Yang-Mills-Chern-Simons theory.

8. There exist other studies in the literature concerning finiteness
of pure Chern-Simons theory \cite{Blasi}. They use a particular
symmetry of the gauge-fixed action in the Landau gauge called `vector
supersymmetry', which has nothing to do with the ordinary
supersymmetry we have discussed. These articles use abstract
cohomology arguments and do not discuss regularization, nor they
include in the classical action a Yang-Mills term. A cohomological
study of $N\!=\!2$ Yang-Mills-Chern-Simons theory in a
nonsupersymmetric gauge has been performed
in ref. \cite{Piguet2}, where only a partial proof of finiteness
exploiting the fact that the theory has extended supersymmetry is
given.

\section*{Acknowledgment}

We thank P. Breitenlohner, J.W. van Holten, F. De Jonghe and D. Maison
for discussions.


\begin{thebibliography}{99}

\bibitem{Piguet} For a superspace approach, see O. Piguet and
   K. Sibold, {\it Renormalized supersymmetry -- The perturbation
   theory of $N=1$ supersymmetric theories in flat spacetime} 
   (Birkhauser, Boston 1986).
   For theories in the Wess-Zumino gauge, without auxiliary fields,
   see P.L. White, Class. Quan. Grav. {\bf 9} (1992) 413; N. Maggiore,
   Int. J. Mod. Phys. {\bf A10} (1995) 3781 and 3937; N. Maggiore,
   O. Piguet and S. Wolf, Nucl. Phys. {\bf B458} (1996) 403
   and {\it Algebraic renormalization of $N=1$ supersymmetric gauge 
   theories with supersymmetry breaking masses}
   {\tt (hep-th/9604002)}. 
\bibitem{tHooft} 
   G. 't Hooft and M.  Veltman, Nucl Phys. {\bf B44} (1972) 189.
\bibitem{GMR} 
   G. Giavarini, C.P. Martin and F. Ruiz Ruiz, Nucl. Phys.  {\bf B381}
   (1992) 222.  
\bibitem{Siegel-DReD} 
   W. Siegel, Phys. Lett.  {\bf 84B} (1979) 193.
\bibitem{Siegel-incon} 
   W. Siegel, Phys. Lett.  {\bf 94B} (1980) 37.
\bibitem{Hepp}
   K. Hepp, {\it Renormalization theory,} in {\it Statistical 
   Mechanics and quantum field theory}, edited by C. DeWitt and R. 
   Stora (Gordon and Breach, New York 1971).\\ 
   H.  Epstein and V. Glasser, Ann. Inst.  Henri Poincar\'e {\bf XIX}
  (1973) 211.  
\bibitem{Maison}
   D. Maison, {\it Renormalization theory, a short account of
   results and problems}, in {\it Renormalization of quantum field
   theories with nonlinear field transformations}, edited by P.
   Breitenlohner, D. Maison and K. Sibold (Springer-Verlag, Berlin 
   1988).
\bibitem{1001} 
   S.J. Gates Jr., M.T. Grisaru, M. Ro\v cek and W.  Siegel, {\it
   Superspace or one thousand and one lessons in supersymmetry}
   (Benjamin, Reading 1983).\\ 
   W. Siegel, Nucl.  Phys. {\bf B156} (1979) 135.
\bibitem{Zinn-Justin} 
   J. Zinn-Justin, {\it Renormalization of gauge theories,} in {\it 
   Trends in elementary particle physics,} edited by H.  Rollnik and 
   K. Dietz (Springer-Verlag, Lectures Notes in Physics 37, 
   Heidelberg 1975).
\bibitem{Breitenlohner} 
   P. Breitenlohner and D. Maison, Commun.  Math. Phys.  {\bf 52} (1977)
   11.  
\bibitem{Bonneau} G. Bonneau, Int.  J. Mod.  Phys. {\bf A}
   (1989). \\ M. Bos, Ann. Phys. {\bf 181} (1988)197.\\ 
   H. Osborn, Ann.  Phys. {\bf 200} (1990) 1.  
\bibitem{Collins} J.C.  Collins, {\it Renormalization}
   (Cambridge University Press, Cambridge 1987). 
\bibitem{Speer} 
   E.R.  Speer, J. Math.  Phys. {\bf 15} (1974) 1; Ann.  Inst. Henri
   Poincar\'e {\bf XXII} (1975).  
\bibitem{Pisarski}
   R. Pisarski and S. Rao, Phys. Rev. {\bf D32} (1985) 2081.
\bibitem{CJvN} 
     D.M. Capper, D.R.T. Jones and P. van Nieuwenhuizen, Nucl. Phys.
     {\bf B167} (1980) 479.
\bibitem{Deser} 
   S. Deser, R. Jackiw and S. Templeton, Ann. Phys. {\bf 140} (1982)
   372. 
\bibitem{Lerda} 
   A. Lerda and P. van Nieuwenhuizen, Phys. Rev. Lett.
   {\bf 62} (1989) 1217.  
\bibitem{Kleppe}
   S. Deser and Z. Yang, Class. Quan. Grav, {\bf 7} (1990) 1603.
   B. Keszthelyi and G. Kleppe, Phys. Lett. {\bf B281} (1992) 33.
\bibitem{Martin}
   C.P. Martin Phys. Lett. {\bf B241} (1990) 513.
\bibitem{GMR2} 
   G. Giavarini, C.P. Martin and F. Ruiz Ruiz, Phys.  Rev. {\bf D47} 
   (1993) 5536; Phys. Lett. {\bf B314} (1993) 328; Phys.  Lett. 
   {\bf B332} (1994) 345.
\bibitem{Schonfeld}
   J. Schonfeld, Nucl. Phys. {\bf B185} (1981) 157.\\
   R. Jackiw and S. Templeton, Phys. Rev. {\bf D23} (1981) 2291.
\bibitem{Blasi}
   A. Blasi and R. Collina, Nucl. Phys. {\bf B345} (1990) 472.\\
   F. Delduc, C. Lucchesi, O. Piguet and S.P. Sorella, Nucl. Phys. 
   {\bf B346} (1990) 313.\\
   C. Lucchesi and O. Piguet, Nucl. Phys. {\bf B381} (1992) 281.
\bibitem{Piguet2}
   N. Maggiore, O. Piguet and M. Ribordy, Helv. Phys. Acta {\bf 68}
   (1995) 264. 
\end{thebibliography}
\end{document}